\documentclass{emulateapj}

\shorttitle{PSR B0656+14 and the Monogem Ring}
\submitted{To appear in the Astrophys. J. Lett.}
\begin{document}

\title{Pulsar PSR B0656+14, the Monogem Ring, and the Origin of the
`Knee' in the Primary Cosmic Ray Spectrum}

\author{
	S. E. Thorsett,\altaffilmark{1}
	R. A. Benjamin,\altaffilmark{2} 
	Walter F. Brisken,\altaffilmark{3} 
	A. Golden,\altaffilmark{4} \&
	W. M. Goss\altaffilmark{3}}
\altaffiltext{1}{Department of Astronomy and Astrophysics, University of
		California, Santa Cruz, CA 95064; \email{thorsett@ucolick.org}}
\altaffiltext{2}{Department of Physics, University of Wisconsin,
Madison, WI 53706; \email{benjamin@physics.wisc.edu}}
\altaffiltext{3}{National Radio Astronomy Observatory, PO Box O, Socorro, 
NM 87801; \email{wbrisken@nrao.edu}, \email{mgoss@nrao.edu}}
\altaffiltext{4}{National University of Ireland, Newcastle Road,
Galway, Republic of Ireland; \email{agolden@it.nuigalway.ie}}

\begin{abstract}
The Monogem ring is a bright, diffuse, $25^\circ$-diameter
supernova remnant easily visible in soft X-ray images of the
sky. Projected within the ring is a young radio pulsar, PSR
B0656+14. An association between the remnant and pulsar has been
considered, but was seemingly ruled out by the direction and magnitude
of the pulsar proper motion and by a distance estimate that placed the
pulsar twice as far from Earth as the remnant. Here we show that in
fact the pulsar was born very close to the center of the expanding
remnant, both in distance and projection. The inferred pulsar and
remnant ages are in good agreement. The conclusion that the pulsar and
remnant were born in the same supernova explosion is nearly
inescapable. The remnant distance and age are in remarkable
concordance with the predictions of a model for the primary cosmic ray
energy spectrum in which the `knee' feature is produced by a single
dominant source.
\end{abstract}


\section{Introduction}

A bright soft X-ray enhancement in the Monoceros and Gemini
constellations was first detected in rocket experiments over thirty
years ago \citep{bckm71}, and resolved into a shell-like structure by
{\it HEAO-2} a decade later \citep{nch+81}. Spectral studies with {\it
ROSAT} confirmed that this so-called `Monogem ring' is a supernova
remnant, probably in the adiabatic expansion phase \citep{psa+96}. The
distance is poorly constrained by evolutionary
arguments---self-consistent Sedov-Taylor models were found by
\citet{psa+96} at all distances between 100 and 1300 pc---but
distances around 300 pc are preferred on grounds of supernova
energetics. At this distance, the model age is 86,000 years, explosion
energy is $1.9\times10^{50}$~erg, and current radius is 66 pc. The
inferred interstellar medium density is
$5\times10^{-3}\mbox{cm}^{-3}$, typical of the hot interstellar
medium.

The radio pulsar PSR B0656+14 lies very close in projection to the
center of the Monogem ring, and an association between the objects
seems natural \citep{nch+81,chmm89,tchf91}. The pulsar is young, with
a characteristic spin-down timescale of 110 kyr \citep{tml93}, pulsed
non-thermal optical and X-ray emission, and unpulsed thermal X-ray
emission consistent with a $10^5$ year-old cooling neutron star
\cite{fok93}. But because the pulsar distance was estimated from
interstellar dispersion measurements \citep{tml93} to be 760 pc---more
than twice the best estimate for the remnant---and a proper motion
measurement appeared to show the pulsar moving towards the center of
the remnant \citep{tc94}, a physical association has been widely
regarded as unlikely \citep[for example]{kas98b}. New very long
baseline interferometric measurements (Brisken et al.\ 2003) lead us
to reconsider this negative conclusion. The distance, now accurately
known from parallax to be 
$288^{+33}_{-27}$~pc, 
is much lower than
previously believed, and the proper motion, 44 mas/yr, is about 40\%
smaller. As we will show, these measurements, together with a
re-examination of the X-ray data, convincingly demonstrate that the
pulsar and remnant were born in a single supernova explosion about
$10^5$ yrs ago.

\section{The Pulsar-Remnant Association}
\begin{figure}
\plotone{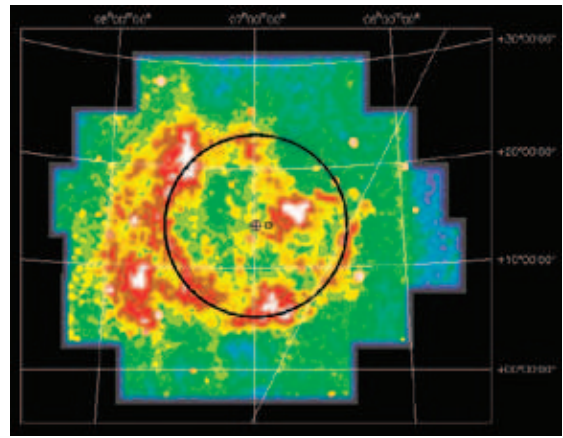}
\caption{The Monogem ring, as seen in the {\it ROSAT} all-sky survey
in the 0.25--0.75~keV x-ray band. PSR B0656+14 is marked with
cross-hairs, and the $9.2^\circ$ circle centered on this point shows
the primary ring structure. The estimated position of the pulsar
$10^5$~yrs ago is marked with a small square. J2000 coordinates are
shown, and the Galactic plane is indicated by a diagonal line. The
ring is imperfect: there is an apparent blow-out to the west, at high
galactic latitude, and a missing quadrant to the northwest, perhaps
due to foreground absorption or to slower expansion into a dense
region. The bright source at 06:17 +22:34 is the SNR IC443.}
\end{figure}
\subsection{Agreement in projection}
The Monogem ring (Fig.~1) shows significant deviations from circular
symmetry, with what may be a blow-out to the west, away from the
Galactic plane, and a missing quadrant in the northwest, where
foreground absorption is relatively high \citep{psa+96}. Because of
the lack of symmetry, the morphological center of the ring is somewhat
uncertain.  For example, at low energy (e.g., the {\it ROSAT} R1 band
image displayed as Fig.~6 in Plucinsky et al.\ 1996), the portion of
the ring in the galactic plane is not readily visible, making the
remnant appear smaller and offset slightly away from the plane; the
arc is evident in higher energy observations (Fig.~1, and also note
Fig.~8 of Plucinsky et al.\ 1996), suggesting the importance of
greater absorption on lines of sight near the plane.

In projection, the pulsar, at $(l,b)\sim(201.1^\circ,+8.3^\circ)$,
lies well within the ring. To test the consistency of its position
with the morphological center of the ring, we have fitted `by eye' a
circle to the incomplete ring, centered on the pulsar. This fit, shown
in Fig.~1, demonstrates that the pulsar's position is not inconsistent
with the morphological center of the ring, within admittedly large
uncertainties.

With the measured proper motion, 44 mas/yr, the pulsar has moved about
$1.25^\circ$ in $10^5$~yrs, from roughly
$(l,b)\sim(200.5^\circ,+7.2^\circ)$. This inferred birth position
appears slightly further from the current center of the ring, but not
unacceptably so.  It is also important to distinguish the
morphological center of the remnant from its true dynamical center: we
should expect an old remnant expanding into a medium with density
decreasing away from the galactic plane to have its apparent center
offset from its explosion center in the direction away from the plane,
just as observed \citep{hp99}.

In the end, we conclude that the pulsar position at birth was within a
few degrees of the current morphological center of the ring. We note
that the {\it a priori} likelihood of an unrelated background source
projected within the ring falling this close to the center is about
5\%. However, on its own the positional coincidence is a strong
argument neither for nor against an association.

\subsection{Agreement in distance}

The strongest argument against a physical association between the
pulsar and remnant has been the pulsar distance, estimated at 760~pc
from the measured column density of free electrons. Sedov modeling
formally allows a wide range of distances for the Monogem ring, but a
distance of about 300~pc has been preferred. Although estimates of
pulsar distances through interstellar dispersion are fairly crude (the
quoted uncertainty was 25\% \citep{tml93}) the distance discrepancy
appeared significant. The small parallax distance,
$288^{+33}_{-27}$~pc, thus came as a surprise. (We note, though, that
the greater distance implied an unacceptably large thermal X-ray
luminosity: an attempt to directly estimate the distance from X-ray
modeling yielded $\sim200$~pc or less \citep{gs99}.)

With this precise pulsar distance available, we consider whether the
distance estimate for the remnant can also be improved. A strong
interstellar O~VI
absorption feature at $+105$~km/s has been detected in the star
15~Mon, in the southern region of the ring \citep{jss98}. The parallax
distance of 15~Mon is $313^{+93}_{-58}$~pc \citep{plk+97}. Because
O~VI at such high velocity is rarely found in the disk, and the
velocity is similar to that expected from a remnant at this stage, we
consider this to be a secure upper limit to the distance of the
Monogem ring. Other evidence suggests the ring is no farther than the
pulsar. The absorption is low,
$N_H\lesssim1.0\times10^{20}\mbox{cm}^{-2}$ \citep{psa+96}, compared
to $\sim1.5\times10^{20}\mbox{cm}^{-2}$ for the pulsar
\citep{esgm00}. The pulsar dispersion measure is much higher than
expected at a distance of 290 pc, suggesting a line of sight through a
highly ionized region.

It is also unlikely that the remnant is significantly closer than the
pulsar. At an assumed distance of 300~pc, the inferred explosion
energy is already low: $1.9\times10^{50}$~erg. A distance below 230~pc
would imply an implausibly low energy, below $10^{50}$~erg. We
conclude that the pulsar and remnant distances agree to within
$\sim50$~pc. Indeed, since the shell radius is $\sim70$~pc, the
evidence is strong that the pulsar is currently within the expanding
supernova shell.

\subsection{Agreement in age}

Finally, we must consider the relative ages of the pulsar and
remnant. The remnant age, from Sedov modeling, has been estimated at
86,000~yrs for a distance of 300~pc.  Two lines of evidence suggest a
comparable age for the neutron star. First, its current temperature is
in good agreement with standard cooling models for a $10^5$~yr old
neutron star \citep{vle95}. Second, the characteristic spin-down
timescale for the pulsar is $P/2\dot P=1.1\times10^5$~yrs.

The second argument deserves more comment.  The age of a pulsar slowing with
a torque proportional to a constant power $n$ of its frequency, $\dot
f\propto -f^n$, can be expressed as
\begin{equation}
\tau=\frac{P}{(n-1)\dot P}\left[1-\left(\frac{P_0}{P}\right)^{n-1}\right],
\end{equation}
where $P_0$ is the initial spin period.  For magnetic-dipole braking,
the ``braking index'' $n=3$, so if the pulsar is born spinning much
faster than its current period, $\tau\approx P/2\dot P$, while the age
is overestimated if the pulsar were born near its current period.

For a handful of very young pulsars, where the braking index can be
measured directly, $n\lesssim3$ \citep{bfg+03}. For middle-aged
pulsars, the good agreement between timing ages and so-called kinetic
ages ($z/\dot z$, where $z$ is the height above the Galactic plane)
requires a mean braking index near 3
\citep{bai89,bwhv92,lbh97}. We acknowledge that the timing age can be 
misleading. Several putative pulsar-SNR
associations suggest large discrepancies between timing and true ages:
for example, PSR J1012$-$5226, in the $\sim7$~kyr old remnant PKS
1209$-$51/52, has a spin-down age of $\sim200-900$~kyr \citep{pzst02},
and PSR J1811$-$1925, with a spin-down age of 24~kyr, has been
associated with the SNR of AD~386 \citep{ttd+99}. Nevertheless, a
recent comparison of kinetic and timing ages for 21 pulsars under
10~Myr in age found a typical discrepancy of only $\sim40\%$
\citep{bfg+03}. We conclude that both the timing and
cooling ages of the pulsar are consistent with the remnant age.

\subsection{Summary of arguments for an association}

The angular position, distance, and age of the pulsar and remnant are
all in excellent agreement with the hypothesis that both were born at
the same location, $\sim300$~pc away, $10^5$~yrs ago. The likelihood
that this is merely a positional and temporal coincidence is extremely
low. Crudely estimating the Galactic Type II supernova rate at one per
century and the region within 300~pc of Earth as
$(300\mbox{\,pc}/10\mbox{\,kpc})^2\approx0.1\%$ of the area of the
Galactic disk, the mean supernova interval in this region is about
100~kyr. The chance that two unrelated supernovae occur within $\sim25$~kyr,
within $5^\circ$ on the sky, and within 50~pc in radial distance is
vanishingly small. Perhaps the most persuasive argument for an
association comes from assuming the opposite. If this $10^5$~yr old
remnant is not that of the supernova that formed the $10^5$~yr old
pulsar, then where is the remnant of that supernova? Many old remnants
are of course invisible because of distance or environment, but if the
Monogem ring and pulsar were formed in different supernovae then they
occurred in very close physical and temporal proximity. The Monogem
ring itself is evidence that a remnant of age $10^5$ years expanding
into this particular environment is visible at 300~pc. We conclude
that a single supernova, 300~pc away and a hundred thousand years ago,
formed both PSR B0656+14 and the Monogem ring.

\section{The Monogem ring as a `single source' for PeV cosmic rays}

Supernova remnants are generally believed to be acceleration sites for
cosmic rays \citep{shk53b,bo80,ber96,bk99}, and in this case it is
intriguing to note that the Monogem ring may hold a clue to a
long-standing puzzle: the origin of the steepening in the primary
cosmic-ray spectrum at about $3\times10^{15}$~eV (3~PeV), called the
`knee.'

Between about $10^{10}$ and a few times $10^{18}$~eV, the differential
cosmic ray energy spectrum is well-described by a broken power law,
proportional to below the knee and above the knee. (A recent review
can be found in \citet{wef03}.) Possible explanations for the knee are
extremely varied, including loss of the most energetic particles from
the Galaxy, unknown physical processes in the development of the
atmospheric shower through which the cosmic rays are detected, or,
most likely, a termination in the acceleration process. A typical
cut-off energy for acceleration in supernova remnants is
$Z\times10^{14}$~eV, where $Z$ is the nuclear charge, though the exact
cut-off will vary by perhaps an order of magnitude with variations in
the explosion energy, magnetic field, interstellar medium density, and
age \citep{lc83,ber96,ew97}.

Recently, Erlykin and Wolfendale (EW) have drawn attention
\citep{ew97} to the sharpness of the knee feature, which has been a
challenge for models in which the break arises from propagation
effects or from a stochastic superposition of multiple sources with
varying high-energy cut-offs. EW propose that around the knee a single
nearby source dominates the cosmic ray spectrum, which is otherwise a
smooth superposition of the contributions from many supernova remnants
throughout the Galaxy and from whatever sources provide the higher
energy cosmic rays, up to $10^{20}$~eV and beyond. At energies of a
few PeV, this single source alone produces 60\% of the flux at
Earth. (This possibility has also recently been considered by other
authors \citep[for example]{bk99}.) Although the data are not yet
conclusive, EW have identified the knee as most likely due to oxygen
nuclei, with a smaller second break at about 10~PeV due to iron
nuclei.

The shape and amplitude of the knee feature have led EW to predict
\citep{ew03} that the single source is a 90--100~kyr old supernova
remnant between 300 and 350~pc from Earth, expanding in an under-dense
medium. The match with the properties of the Monogem ring is striking,
though almost certainly in part coincidental considering the large
remaining uncertainties in cosmic ray acceleration and diffusion
models. Nevertheless, an important objection to the single source
model for the knee feature has been the lack of a suitable source
\citep{bha02}.  That objection now appears to have been removed.

\acknowledgements

S.E.T. is supported by
the NSF under grant AST-0098343,  R.A.B. is supported by NASA ATP
grant NAG5-12128, and A.G. is supported by Enterprise Ireland 
under grant SC/2001/322. This research has made use of the {\it ROSAT}
all-sky survey data, which have been processed at MPE, and of the
telescopes of the National Radio Astronomy Observatory, a facility of
the NSF operated under cooperative agreement by Associated
Universities, Inc. We thank an anonymous referee for suggestions that
improved the manuscript.


\end{document}